\documentstyle[11pt,aaspp4]{article}
\begin{document}

\large \bf A giant, periodic flare from the soft gamma repeater SGR1900+14

\vspace{.25in}

\rm \normalsize
K. Hurley$^a$, T. Cline$^b$, E. Mazets$^c$, S. Barthelmy$^b$, P. Butterworth$^b$,  F. Marshall$^b$ , D. 
Palmer$^b$ R. Aptekar$^c$, S. Golenetskii$^c$, V. Il'Inskii$^c$, D.  Frederiks$^c$,  J. McTiernan$^a$, R. 
Gold$^d$, J. Trombka$^b$

\vspace{.25in}

a UC Berkeley Space Sciences Laboratory, Berkeley, CA 94720-7450

b NASA Goddard Space Flight Center, Code 661, Greenbelt, MD 20771

c Ioffe Physical-Technical Institute, St. Petersburg, 194021, Russia

d The Johns Hopkins University, Applied Physics Laboratory, Laurel, MD 20723

\vspace{.25in}

\bf Soft gamma repeaters are high-energy transient sources  associated with neutron 
stars in young supernova remnants$^1$.  They emit sporadic, short ($\sim$ 0.1 s) bursts with 
soft energy spectra during periods of intense activity.  The event of March 5, 1979 
was the most intense and the only clearly periodic one to date$^{2,7}$.  Here we report on 
an even more intense burst on August 27, 1998, from a different soft gamma 
repeater, which displayed a hard energy spectrum at its peak, and was followed by a 
$\sim$ 300 s long tail with a soft energy spectrum and a dramatic 5.16 s period.  Its peak 
and time integrated energy fluxes at Earth are the largest yet observed from any 
cosmic source.  This event was probably initiated by a massive disruption of the 
neutron star crust,  followed by an outflow of energetic particles rotating with the 
period of the star.  Comparison of these two bursts supports the idea that magnetic 
energy plays an important role, and that such giant flares, while rare, are not 
unique, and may occur at any time in the neutron star's activity cycle.

\rm Four soft gamma repeaters (SGRs) are known. All appear to be associated with radio 
supernova remnants, indicating that they are young$^4$ ($<$20,000 y).  SGRs are probably 
strongly magnetized neutron stars ('magnetars'$^5$), in which, unlike the radio pulsars, the 
magnetic energy dominates the rotational energy.  SGR0525-66 produced both the 
unusual, energetic and periodic burst of March 5 1979$^{6,7,8}$ and a series of subsequent, 
much smaller bursts$^{9,10}$.  It lies towards the N49 supernova remnant in the Large 
Magellanic Cloud$^{11,12}$. A quiescent soft X-ray source has been identified which may be 
the neutron star$^{13}$.  SGR1900+14, first detected in 1979, was, until recently, the least 
prolific SGR$^{14,15}$, hindering attempts to locate it precisely.  Several lines of evidence 
suggested that it was associated with the galactic supernova remnant G42.8+0.6$^{16}$ and a 
quiescent soft X-ray source$^{17}$.  This possible association was strengthened by a source 
location obtained with the network synthesis method$^{18}$, and more recently by 
triangulation$^{19,20,21}$, although since this X-ray source lies outside the remnant, the 
connection between the two could still be considered to be unresolved.

An observation of the quiescent soft X-ray source possibly associated with SGR1900+14 
by the ASCA spacecraft in April 1998 showed that the X-rays exhibited a 5.16 s period$^{22}$.  
In May, SGR1900+14 came out of a long dormant phase, emitting strong, frequent 
bursts$^{19,23}$.  On August 27, it emitted the exceptionally intense giant flare reported here, 
detected by instruments on GGS-Wind$^{24}$, Ulysses$^{21}$, the Rossi X-Ray Timing Explorer$^{25}$ 
(RXTE), BeppoSAX, and the Near Earth Asteroid Rendezvous (NEAR). The entire event 
profile is shown in figure 1 with Ulysses 0.5 s resolution data.  In very general terms, the 
burst rose to a maximum and decayed roughly as a power law in time with an index of $\sim$-
1.8.  However, the event onset is complex; Konus-Wind observations resolve components 
$<$4 ms. A sinusoidal component dramatically modulated the later part of the profile for 
the duration of the observation with varying amplitudes, the first direct detection of the 
5.16 s periodicity at hard X-ray energies.  The inset to Figure 1 shows 31.25 ms time 
resolution Ulysses data, demonstrating that the 5.16 s pulsations commenced 
approximately 35 s after the peak. It is clear that the pulse profile is considerably more 
complex than a single sinusoidal curve, with at least 4 maxima and minima in a single 
cycle.

A remarkable coincidence, the initiation of NEAR gamma-ray monitoring only days 
before August 27th but after many months of silent cruise towards Eros, made possible 
the high-precision source localization of this event by triangulation, i.e. analysis of the 
arrival times at Ulysses, GGS-Wind, RXTE, and NEAR.  This is the only time, other than 
for the March 5, 1979 event$^{11,12}$, that an SGR has been localized by triangulation at three 
or more widely separated spacecraft, leading directly to an error box.  All six source 
annuli, determined from the various two-spacecraft comparisons, are consistent with the 
coordinates of the quiescent soft X-ray source$^{18,21}$ (RA(J2000) =  
19 h 07 m 14 s, Dec(J2000) = 9$^{\rm o}$ 19' 19"). The details will be reported elsewhere, but we note that this positional 
agreement, as well as the agreement between the periodicities found in soft X-rays and in 
the giant flare light curve, now leave no doubt about the association between the SGR 
and the quiescent X-ray source.

The temperature of the energy spectrum of this event is shown in figure 1.  With the 
exception of the peak, the temperature is kT$\sim$30 keV, which is similar to SGR bursts in 
general.  At the peak, however, the temperature averaged over a 1 s interval is kT $\sim$ 240 
keV.  Finer time resolution measurements were recorded by Konus, indicating a  peak 
temperature $\sim$ 1200 keV, and a maximum photon energy of 2 MeV.  Hard spectra such as 
these are not characteristic of SGR bursts;  one was observed for the peak of the March 5 
1979 event$^{6,26}$. Table 1 compares the properties of these two giant flares.
Comparisons between very intense bursts observed by different instruments are subject to 
numerous uncertainties.  Dead time effects, different time resolutions and energy ranges, 
and pulse pile-up are difficult or even impossible to correct for; hence the "approximate" 
and "greater than" symbols in Table 1.  However, to within these uncertainties, the 
parameters of the August 27 1998 event are consistent with it having the largest peak flux 
and fluence of any of the several thousand SGRs and cosmic gamma-ray bursts observed 
to date.

Recently it has been suggested$^{23,27}$ that the neutron stars associated with SGRs are 
magnetars, i.e. that they have magnetic fields of several times 10$^{14}$ G$^5$.  This is based on 
observations of the quiescent counterparts in X-rays, which display pulsations with a 
slowly lengthening period; the spin-down is interpreted as due to magnetic dipole 
radiation.  In the magnetar model, the giant flares of August 27 and March 5 are due to a 
readjustment of the magnetic field, accompanied by a massive, large-scale cracking of the 
neutron star crust.  In both cases the initial hard spectrum would be produced by the 
conversion of magnetic energy to energy in a clean electron-positron and photon fireball 
uncontaminated by ions, which would soften the spectrum.  The highest energy photons 
observed are only slightly above the electron-positron pair production threshold; this is 
consistent with attenuation due to this process, although there is at present no direct 
evidence for a cutoff.  Expanding away from the stellar surface, part of the fireball would 
be trapped in the magnetosphere, producing the observed soft tails.  The periodicity 
indicates that this emission was either anisotropic and/or that it occurred close enough to 
the neutron star to be occulted by it; the decay in intensity with approximately constant 
spectral temperature is interpreted as a shrinking in the volume of the emission region.  
The complex pulse structure implies that several regions of the magnetosphere were 
involved.  It is noteworthy that, despite the factor of 25 difference between the peak 
luminosities of the August 27 and March 5 events, the ratios of peak to total energy are 
within a factor of 2 of each other, suggesting that similar magnetic field geometries may 
play an important role.  Since the soft spectrum which follows the intense main peak in 
both cases is attributed to radiation from an optically thick pair plasma trapped in the 
neutron star's magnetosphere, the magnetic field strength may be estimated from the 
energy in this component$^5$:

$
B > 4X10^{14} \left( \frac{\Delta R}{10 km} \right) ^{-3/2} \left( \frac{1+ \Delta R/R}{2} \right) ^3 \left( \frac {E_{tail}}{3.6x10^{44} erg} \right) ^{1/2} G
$

Where R is the radius of the neutron star and $\Delta$ R ($\sim$10 km) is the outer radius of the 
magnetic flux loop containing the pair plasma.  For the March 5 event, this gives 
$B>4x10^{14} G$; for August 27, $B>10^{14} G$,  providing a confirmation of the magnetar model 
which is independent of the observation and interpretation of the spin-down, but 
consistent with it.

The existence of a strong magnetic field helps to explain the high luminosities 
encountered in both events, five to six orders of magnitude greater than the Eddington 
limit.  A strong magnetic field suppresses the Compton scattering cross-section, and 
reduces the opacity$^5$.

The giant flare of March 5 1979 was observed to precede the much smaller event series 
from SGR0525-66.  Observations during the preceding six months failed to reveal any 
source activity, and it was speculated at the time that this was a unique, catastrophic 
event in the life of a neutron star, and one that initiated the series of bursts subsequently 
observed.  Our observation of the August 27 1998 event leads to a different 
interpretation.  The source evolved from a weak, infrequent repeater to an intensely 
active one, indicating that the neutron star's crust was able to adjust to magnetic stresses 
by undergoing relatively minor, localized cracking for a long period.  The small precursor 
to the giant flare was comparable in intensity to these bursts, and may have been the final 
trigger for it.  In the following months, these bursts have continued. Thus our 
observations imply that rare giant flares on SGRs may be the rule, rather than the 
exception, and that they may occur at any time.  It therefore seems likely that SGR0525-
66 emitted relatively weak bursts prior to March 5, 1979, which went undetected due to 
spacecraft coverage and/or weakness. The magnetar theory predicts that on any given 
SGR, such events may recur on a timescale of $\sim$decades or more$^{28}$.  It is now almost two 
decades since the March 5 event; future monitoring of this and other SGRs can confirm 
this idea.

\vspace{0.25in}
\bf References

\rm 
1. Hurley, K., Are the Soft Gamma Repeaters a Motley Crew? , in 3rd Huntsville 
Symposium, AIP Conf. Proc. 384 (AIP: New York), Eds. C. Kouveliotou, M. 
Briggs, and G. Fishman, 889-896 (1996)

2. Mazets, E., Golenetskii, S., Guryan, Yu, and Ilyinskii, V., The 5 March 1979 event and 
the distinct class of short gamma bursts: are they of the same origin?, Astrophys. 
Space Sci. 84, 173-189 (1982)

3. Cline, T., The Unique Cosmic Event of 1979 March 5, Comments on Astrophysics 
1,13-20 (1980)

4. Braun, R., Goss, W., and Lyne, A., Three Fields Containing Young Pulsars: The 
Observable Lifetime of Supernova Remnants, Ap. J. 340, 355-361 (1989)

5. Thompson, C., and Duncan, R., The Soft Gamma Repeaters as Very Strongly 
Magnetized Neutron Stars. I. Radiative Mechanism for Outbursts, Mon. Not. R. Astron. 
Soc. 275, 255-300 (1995)

6. Mazets, E., et al., T., A Flaring X-Ray Pulsar in Dorado, Nature, 282, 587-589 (1979)

7. Barat, C., et al., Evidence for Periodicity in a Gamma Ray Burst, Astronomy and 
Astrophysics Lett., 79, L24-L25 (1979)

8. Cline, T., et al., Detection of a Fast, Intense and Unusual Gamma Ray Transient, Ap. J. 
Lett. 237, L1-L5 (1980)

9. Golenetskii, S., Ilinskii, V., Mazets, E., and Guryan, Yu., Repeating Gamma Ray 
Bursts From the Source FXP 0520-66, Sov. Astron. Lett., 5, 340-342 (1979)

10. Golenetskii, S., Ilyinskii, V., Mazets, E., Recurrent Bursts in the source of the 5 
March 1979 event, Nature 307, p. 41-43 (1984)

11. Evans, W., et al., Location of The Gamma-Ray Transient Event of 1979 March 5, Ap. 
J. Lett. 237, L7-L9 (1980)

12. Cline, T., et al., Precise Source Location of the Anomalous 1979 March 5 Gamma 
Ray Transient, Ap. J. Lett., 255, L45-L48 (1982)

13. Rothschild, R., Kulkarni, S., and Lingenfelter, R., Discovery of an X-ray Source 
Coincident with the Soft Gamma Repeater 0525-66, Nature 368, 432-434 (1994)

14. Mazets, E., Golenetskii, S., and Guryan, Yu., Soft Gamma Ray Bursts From the 
Source B1900+14, Sov. Astron. Lett., 5(6), 343-344 (1979)

15. Kouveliotou, C., et al., Recurrent Burst Activity from the Soft Gamma-Ray Repeater 
SGR 1900+14, Nature 362, 728-730 (1993)

16. Kouveliotou, C., et al., The Rarity of Soft Gamma-Ray Repeaters Deduced from 
Reactivation of SGR1806-20, Nature 368, 125-127 (1994)

17. Vasisht, G., Kulkarni, S., Frail, D., and Greiner, J., Supernova Remnant Candidates 
for the Soft Gamma-Ray Repeater 1900+14, Ap. J. 431, L35-L38 (1994)

18. Hurley, K., et al., Network Synthesis Localization of Two Soft Gamma Repeaters, 
Ap. J. 431, L31-L34 (1994)

19. Hurley, K., Kouveliotou, C., Mazets, E., and Cline, T., SGR1900+14, IAUC 6929, 
(1998a)

20. Hurley, K. et al., Reactivation and Precise IPN Localization of the Soft Gamma 
Repeater SGR1900+14, ApJ, in press (1998b) 

21. Hurley, K., Cline, T., Mazets, E., and Golenetskii, S., SGR1900+14, IAUC 7004,  
(1998c)

22. Hurley, K., Kouveliotou, C., Murakami, T., and Strohmayer, T., SGR1900+14, IAUC 
7001 (1998e)

23. Kouveliotou, C., Strohmayer, T., Hurley, K., van Paradijs, J., and Woods, P., 
SGR1900+14, IAUC 7001 (1998b)

24. Cline, T., Mazets, E., and Golenetskii, S., SGR1900+14, IAUC 7002 (1998)

25. Marshall, F., et al., IAUC 7005 (1998)

26.  Fenimore, E., Klebesadel, R., and Laros, J., The 1979 March 5 Gamma-Ray 
Transient: Was It a Classical Gamma-Ray Burst?, Ap. J. 460, 964-975 (1996)

27. Kouveliotou, C., et al., An X-ray Pulsar With a Superstrong Magnetic Field in the 
Soft Gamma-Ray Repeater SGR1806-20, Nature, 393, 235-237 (1998)

28. Duncan, R., private communication (1998)

29. Hurley, K., et al., The Solar X-ray/Cosmic Gamma-ray Burst Experiment Aboard 
Ulysses, Astron. Astrophys. Suppl. Ser. 92, 401-410 (1992)

30. Barat, C. et al., Fine Time Structure in the 1979 March 5 Gamma-Ray Burst, Astron. 
Astrophys. 126, 400-402, 1983

\bf Acknowledgments.  \rm We are grateful to R. Duncan for discussions about the magnetic 
field estimate.  This work was supported in the U.S. by a grant from NASA and a 
contract from JPL, and at the Ioffe Institute by an RSA contract.

\begin{tabular}{|p{1.5in}|p{1.5in}|p{1.5in}|} \hline
\multicolumn{3}{|c|}{\bf Table 1.  Properties of the August 27 1998 and March 5 1979 bursts} \\ \hline 
                            & August 27 1998              & March 5 1979 \\ \cline{1-3}
Rise time                  & Complex, structures $<$4 ms & Simple, $<$2 ms  \\ \cline{1-3}
Morphology of main peak    & Complex structure, duration $\sim$1 s & Complex structure, duration 
$\sim$150 ms$^{30}$ \\ \cline{1-3}
Periodicity                & 5.16 s &  8.1 s \\ \cline{1-3}
Peak flux, erg cm$^{-2}$ s$^{-1}$ & $\geq$ 3.4x10$^{-3}$ , $>$ 25 keV & $\sim$1.5x10$^{-3}$ , $>$ 50 keV \\ \cline{1-3}
Fluence, erg cm$^{-2}$ &             $\geq$ 7x10$^{-3}$   & $\sim$2x10$^{-3}$ \\ \cline{1-3}
Spectrum at peak, kT (keV)   & 240 (average over 1 s) & 246 (average over 200 ms)$^{26}$ \\ \cline{1-3}
Highest photon energy in peak & 2 MeV & $>$ 1 MeV \\ \cline{1-3}
Spectrum of pulsations, kT (keV) & 30 & 30 \\ \cline{1-3}
Source distance, kpc & $\sim$7 (G42.8+0.6) & $\sim$50 (N49) \\ \cline{1-3}
Peak source luminosity, erg/s & $\geq$ 2x10$^{43}$ & $\sim$5x10$^{44}$ \\ \cline{1-3}
Precursor observed? & Yes & No \\ \cline{1-3}
Delay between main peak and periodic emission & 35 s & None \\ \cline{1-3}
Ratio of energy in main peak to total energy in burst & 0.46 & 0.25 \\ \cline{1-3}
Source activity in months preceding the burst & Intense & None observed \\ \cline{1-3}

\end{tabular}

\newpage

\begin{figure}
\plotone{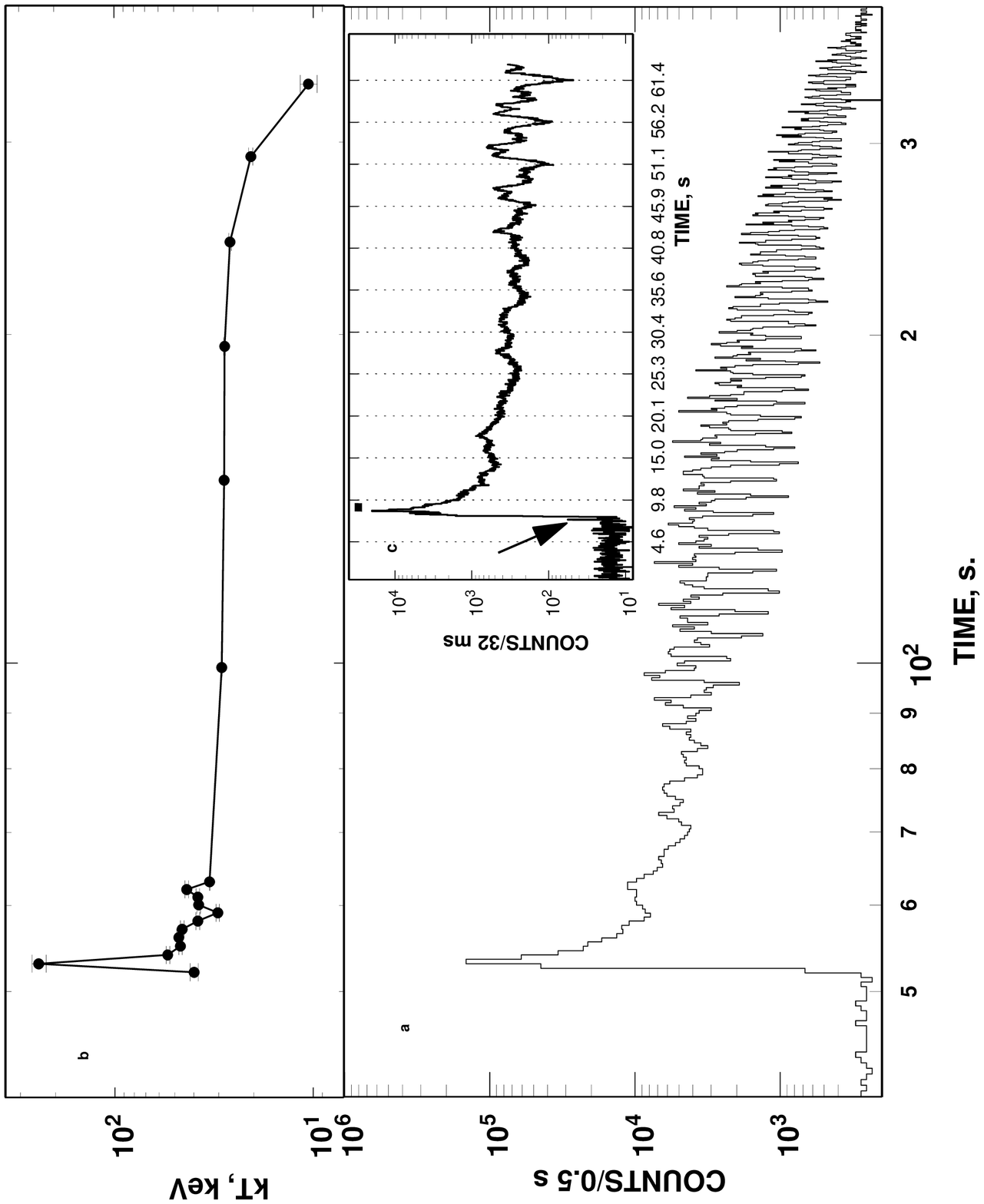}

\end{figure}

Figure 1. Ulysses data for the August 27 1998 giant flare.

a.  25-150 keV time history, corrected for dead time effects, from the 0.5 s resolution 
continuously available real time data.  Zero seconds corresponds to 37283.12 s UT at 
Earth. This event was so intense that it temporarily saturated or shut down some 
experiments, but because of the relatively small detection area of the Ulysses$^{29}$ sensor (20 
cm$^2$) it was not subject to severe dead time or pulse pile-up problems; in fact solar flare 
data producing considerably higher count rates have been successfully analyzed with this 
instrument.

b. Spectral temperature as a function of time.  The spectra were measured by Ulysses in 
intervals with increasing durations of 1 - 48 s.  No simple, two-parameter fit describes the 
spectrum well, in part because the measurement uncertainties are dominated by 
systematic effects.  However, we have used an optically thin thermal bremsstrahlung 
spectrum to characterize grossly the spectral temperature.

c. 0.03125 s resolution time history of the event from the triggered data, available for 64 
s.  The burst triggered on the precursor (arrow) $\sim$0.4 s prior to the main peak.  A grid is 
drawn to indicate the 5.16 s periodicity, showing its absence for the first $\sim$35 s after the 
main peak.  The short horizontal line at the top indicates the position of the hard spectral 
peak measured by Ulysses.  Zero seconds corresponds to 37327.81 s UT at Earth.
\end{document}